\input{style/arxiv-ba.cfg}
\documentclass[ba,linksfromyear,dvips,preprint]{imsart}

\usepackage{amsthm,amsmath}
\usepackage{amssymb}
\usepackage{graphicx}

\usepackage{multirow}
\usepackage{color,colortbl}
\usepackage{graphicx}
\definecolor{Gray}{gray}{0.9}

\pubyear{2015}
\volume{10}
\issue{1}
\firstpage{53}
\lastpage{74}
\doi{10.1214/14-BA880}% Updated by VTEXPTS2LaTeX.exe, 21.01.2015 15:23

\begin{document}

\begin{frontmatter}
\title{Bayesian Polynomial Regression Models to Fit Multiple Genetic
Models for Quantitative Traits}
\runtitle{Polynomial Coding of Genetic Data}

\begin{aug}
\author[a]{\fnms{Harold} \snm{Bae}\corref{}\ead[label=e1]{baeh@bu.edu}},
\author[b]{\fnms{Thomas} \snm{Perls}\ead[label=e2]{thperls@bu.edu}},
\author[c]{\fnms{Martin} \snm{Steinberg}\ead[label=e3]{mhsteinb@bu.edu}},
\and
\author[a]{\fnms{Paola} \snm{Sebastiani}\ead[label=e4]{sebas@bu.edu}}

\runauthor{H. Bae, T. Perls, M. Steinberg, and P. Sebastiani}

\address[a]{Department of Biostatistics, Boston University School of
Public Health, \printead{e1}}
\address[b]{New England Centenarian Study, Section of Geriatrics,
Department of Medicine, Boston University School of Medicine, \printead{e2}}
\address[c]{Center of Excellence in Sickle Cell Disease Boston Medical
Center, Pediatrics, Pathology and Laboratory Medicine, \printead{e3}}
\end{aug}

%% Abstract %%
%
\begin{abstract}
We present a coherent Bayesian framework for selection of the most
likely model from the five genetic models (genotypic, additive,
dominant, co-dominant, and recessive) commonly used in genetic
association studies. The approach uses a polynomial parameterization of
genetic data to simultaneously fit the five models and save
computations. We provide a closed-form expression of the marginal
likelihood for normally distributed data, and evaluate the performance
of the proposed method and existing method through simulated and real
genome-wide data sets.
\end{abstract}

%% Keywords %%
%

\end{frontmatter}

%% Mainmatter %%

%s1 ###
\section{Introduction}
Genome-wide association studies have been a popular approach to
discover genetic variants that are associated with increased risk for
rare and common diseases (\citet{sebas2009}). The most common variants
in the human genome are single nucleotide polymorphisms (SNPs): DNA
bases that can vary across individuals. Typically SNPs have two
alleles, say A and B, and based on the combination of SNPs alleles in
each chromosome pair (the genotype), an individual can be homozygous
for the A allele if both chromosomes carry the allele A, homozygous for
the allele B if both chromosomes carry the B allele, and heterozygous
when the two chromosomes carry the A and B alleles. Genotyping DNA was
a slow and expensive process until mid-2000, when high throughput
technologies produced microarrays that can generate the genetic
profiles of an individual in hundreds of thousands to millions of SNPs,
and the technology was the trigger for an explosion of genome-wide
association studies (GWAS) to discover the genetic base of common diseases.

Typically in a GWAS the association between each SNP and a quantitative
trait is tested using linear regression under a specific genetic model
that can assume a genotypic (2 degrees of freedom), dominant,
recessive, co-dominant, or additive mode of inheritance of each tested
SNP. In a genotypic model the 3 genotypes AA, AB and BB are treated as
a factor with 3 levels. The other 4 genetic models compress the 3
genotypes into a numerical variable by either counting the number of
minor alleles (additive model), or by recoding the genotypes as AA=0
versus AB, BB=1 (dominant model for the B allele), AA, AB=0 versus BB=1
(recessive model for the B allele), AA, BB=0 versus AB=1 (co-dominant
model). However, the inheritance pattern is rarely known, and using a
suboptimal model can lead to a loss of power (\citet{lettre}).

Selecting the correct genetic model for each SNP is often accomplished
by fitting the five models and choosing the model that describes the
data best. This approach has several drawbacks. It increases
computational burden with genome-wide data as 5 GWASs need to be
conducted. Furthermore, testing five models for each SNP increases the
burden of multiple testing in addition to the existing issue of
multiple comparisons with millions of SNPs. More importantly, the
optimal method for choosing the best model is not clear (\citet
{lettre}). The common practice is to simply use the additive genetic
model. It has been shown that additive models perform reasonably well
to detect variants that have additive or dominant inheritance pattern,
but they are underpowered when the true mode of inheritance is
recessive (\citet{bush}). Others (\citet{frei,gonzales,li,so}) have
proposed to study the maximum of the three test statistics derived
under additive, dominant, and recessive models.

We propose a polynomial parameterization of the genetic data that
includes the five genetic models as special cases, and we describe a
coherent Bayesian framework to select the most likely genetic model
given the data. This polynomial parameterization is equivalent to the
genetic model described in \citet{servin} that adds a dominance effect
to the additive model to describe non-additive genetic effects. The
advantage of either parameterization is that, in a Bayesian framework,
fitting a single model becomes sufficient to test the
genotype-phenotype associations without specifying a particular genetic
model and this problem has been described in detail in \citet{servin}.
Here, we focus on the specific task of selection of the best genetic
model when the specific mode of inheritance is of interest in addition
to whether a SNP is associated with the trait.

The next section describes this parameterization and shows that there
is a mathematical relationship between the parameters of the polynomial
model and each of the five possible genetic models. Section 3 describes
the model selection approach that is based on the computation of the
marginal likelihood of the five models so that the model with maximum
posterior probability can be identified. Section 3 also provides closed
form solutions for the marginal likelihood and for the estimates of the
parameters of the model with the highest marginal likelihood or Bayes
Factor (BF), assuming exchangeable observations that follow normal
distributions. The proposed method is evaluated through simulation
studies in Section 4, and is applied to two GWAS data sets in Section
5. Conclusions and suggestions for further work are provided in Section~6.

%s2 ###
\section{Relationship Between the Polynomial Model and Other Genetic Models}
Here we show that the five genetic models are specific cases of a
general polynomial model, with parameters that satisfy some linear
constraints. Let y denote the response variable in the genetic
association study, and consider the polynomial regression model
\[
E(y|\beta)=\beta_0+\beta_1x_{add}+\beta_2x^2_{add}
\]
where $\beta$ denotes the vector of regression parameters and $x_{add}$
is the variable that codes for the genotype data as follows:
\begin{displaymath}
x _{add}= \left\{
\begin{array}{ll}
0& \mbox{if genotype is AA} \\
1 & \mbox{if genotype is AB}\\
2 & \mbox{if genotype is BB}.
\end{array}
\right.
\end{displaymath}
Note that the proposed model is equivalent to the additive model with
dominance effects described in \citet{servin}:
\[
E(y|\theta)=\theta_0+\theta_1x_{add}+\theta_2x_{het}
\]
where $x_{het}=1$ for heterozygous genotype and 0 otherwise, and $\theta
_1=\beta+2\beta_2;$ $\theta_2=-\beta_2$. Mathematically, we found the
polynomial parameterization more appealing as it allows interpretation
of the regression coefficients in terms of the SNP dosage.

\noindent
%s2.1 ###
\subsection{Genotypic Association Model}
The genotypic association model is typically parameterized using two
indicator variables to describe the effect of the genotypes AB and BB
relative to AA:
\[
E(y|\gamma)= \gamma_0+ \gamma_1 x_{AB}+ \gamma_2x_{BB}
\]
\[
x_{AB}=1\ \mbox{if genotype is AB (and 0 otherwise)}
\]
\[
\mbox{and}\ x_{BB}=1\ \mbox{if genotype is BB (and 0 otherwise)}.
\]
This parameterization specifies the expected value of $y$, for each of
the 3 genotypes AA, AB, BB as summarized in Table 1. Equating the
expected values of $y$ from the 2 different parameterizations produces
a system of linear equations:

\[
\left[
\begin{array}{c}
\gamma_0 \\
\gamma_0+\gamma_1 \\
\gamma_0+\gamma_2
\end{array}
\right]=
\left[
\begin{array}{c}
\beta_0 \\
\beta_0+\beta_1+\beta_2 \\
\beta_0+2\beta_1+4\beta_2
\end{array}
\right]
\]
that can be solved as
\[
\left[
\begin{array}{c}
\gamma_0 \\
\gamma_1 \\
\gamma_2
\end{array}
\right]=
\left[
\begin{array}{ccc}
1 & 0 & 0 \\
0 & 1 & 1 \\
0 & 2 & 4
\end{array}
\right]
\left[
\begin{array}{c}
\beta_0 \\
\beta_1 \\
\beta_2
\end{array}
\right]=
\left[
\begin{array}{ccc}
1 & 0 & 0 \\
0 & 1 & 1 \\
0 & 2 & 4
\end{array}
\right]\beta.
\]
Therefore, the parameters in the polynomial model and the genotypic
association model have a one-to-one relationship. For the other genetic
models, some constraints on parameters of the polynomial model are necessary.

\noindent
%s2.2 ###
\subsection{Additive Model}
The parameterization of the additive genetic model is $E(y|\alpha_A)=
\alpha_{A0}+ \alpha_A x_{add}$ and equating the expected values of $y$
in Table 1 leads to the system of linear equations:
\[
\beta_0=\alpha_{A0}
\]
\[
\beta_1+\beta_2=\alpha_{A}
\]
\[
2\beta_1+4\beta_2=2\alpha_{A}
\]
that can be solved if $\beta_2=0$, so that $\beta_0= \alpha_{A0}$, and
$\beta_1=\alpha_A$. Therefore the relationship between the parameters
in the polynomial model and additive model requires a linear constraint
on the vector $\beta$.

\noindent
%s2.3 ###
\subsection{Dominant Model}
Now, consider the dominant model for the B allele: $E(y|\alpha_D)=
\alpha_{D0}+ \alpha_D x_{Dom}$, where $x_{Dom}$=1 if genotype is AB or
BB (and 0 otherwise). Proceeding as in the previous cases leads to the
equations:
\[
\alpha_{D0}=\beta_0
\]
\[
\alpha_D=\beta_1+\beta_2=2\beta_1+4\beta_2 .
\]
The system has the solution $\alpha_{D0}=\beta_0$ and $\alpha_D=\frac
{2}{3}\beta_1$ if the parameters of the polynomial model satisfy the
constraint $\beta_1+3\beta_2=0$.

\noindent
%s2.4 ###
\subsection{Recessive Model}
Similarly, consider the recessive model for the B allele: $E(y|\alpha
_R)= \alpha_{R0}+ \alpha_R x_{Rec}$, where $x_{Rec}$=1 if genotype is
BB (and 0 otherwise). In this case, the relations between parameters are:
\[
\alpha_{R0}=\beta_0=\beta_0+\beta_1+\beta_2
\]
\[
\alpha_R=2\beta_1+4\beta_2
\]
with linear constraint $\beta_1+\beta_2=0$, $\alpha_{R0}=\beta_0$ and
$\alpha_R=2\beta_1$.

\noindent
%s2.5 ###
\subsection{Co-dominant Model}
Lastly, consider the co-dominant genetic model: $E(y|\alpha_C)= \alpha
_{C0}+ \alpha_C x_{Cod}$, where $x_{Cod}$=1 if genotype is AB (and 0
otherwise). In this case:
\[
\alpha_{C0}=\beta_0=\beta_0+2\beta_1+4\beta_2
\]
\[
\alpha_C=\beta_1+\beta_2 .
\]
The linear constraint is $\beta_1+2\beta_2=0$, so that $\alpha
_{C0}=\beta_0$ and $\alpha_C=\frac{1}{2}\beta_1$.

% matrix: b file: table1.tex 14 Sep 2001 18:06:44
%
%t1 ###
\begin{table}[htbp]
\footnotesize
\label{table1}
\begin{center}
\begin{tabular}{|p{1.3cm}|p{1.8cm}|p{1.3cm}|p{1.4cm}|p{1.3cm}|p{1.3cm}|p{1.3cm}|}\hline
& Polynomial Model & 2-df General Model & Additive Model & Dominant
Model & Recessive Model & Co-dominant Model \\ \hline
$E(y|AA)$ & $\beta_0$ & $\gamma_0$ & $\alpha_{A0}$ & $\alpha_{D0}$
&$\alpha_{R0}$ & $\alpha_{C0}$ \\ \hline
$E(y|AB)$ & $\beta_0+\beta_1+\beta_2$ & $\gamma_0+\gamma_1$ & $\alpha
_{A0} + \alpha_A$& $\alpha_{D0} + \alpha_D$ & $\alpha_{R0}$ &$\alpha
_{C0} + \alpha_C$ \\ \hline
$E(y|BB)$ & $\beta_0+2\beta_1+4\beta_2$ & $\gamma_0+\gamma_2$ & $\alpha
_{A0} + 2\alpha_A$ &$\alpha_{D0} + \alpha_D$ &$\alpha_{R0} + \alpha_R$
& $\alpha_{C0} $ \\ \hline
\end{tabular}
\caption{Expected value of the quantitative trait for 3 genotypes in
each model.}
\end{center}
\end{table}

In summary, there is a one-to-one transformation between the parameters
of the polynomial and general model, while the transformation between
the polynomial and the other four models (additive, dominant,
recessive, and co-dominant) is constrained by a linear contrast of the
parameters in the polynomial model.

\noindent
%s3 ###
\section{Model Selection via Marginal Likelihood and Parameter Estimation}
The polynomial parameterization provides a framework to simultaneously
fit different genetic models. Given a sample of genotype data, the
question is how to select the most appropriate genetic model. We
propose a Bayesian model selection approach in which genetic models are
compared based on their marginal likelihood and the model with largest
marginal likelihood is selected, assuming that a priori the 5 genetic
models are equally likely.

In the polynomial model ($y|\beta=X\beta+\epsilon$ in matrix form), the
data are assumed to be exchangeable and follow a normal distribution with:
\[
y|X, \beta, \tau\sim N(X\beta, \frac{1}{\tau}I)
\]
where $I$ is the identity matrix. A standard normal-gamma prior for the
vector of parameters $\beta$ and precision $\tau$ is assumed such that
$p(\beta, \tau)=p(\beta|\tau)p(\tau)$, where
\[
\tau\sim Gamma( a_1, a_2)
\]
\[
\beta|\tau\sim N(\beta_0, (\tau R_0)^{-1})
\]
with $\beta_0$, $R_0$, $a_1$, and $a_2$ as prior hyperparameters.
Specification of these prior hyper-parameters can be subjective and
represents the prior probability of alternative genetic models. With
genome-wide data, most of the tested SNPs are likely to be null SNPs
and it is both reasonable and convenient to assume non-informative
priors. Therefore the following values for the prior hyper-parameters:
$\beta_0=0$, $R_0=I$, $a_1=1$, and $a_2=1$ will be assumed. If there is
strong prior belief about certain genetic models, more informative
prior distributions can be chosen and this problem is described at
length in \citet{servin}. The marginal likelihood given this polynomial
model $M_p$ can be computed analytically in the equation below:
\[
p(y|M_p)= \int p(y|X, \beta, \tau) p(\beta, \tau) d\beta d\tau=
\frac{1}{ (2\pi)^{\frac{n}{2}}} \frac{ a_{2n}^{a_{1n}}\Gamma
(a_{1n})}{a_2^{a_1}\Gamma(a_1) } \frac{ |R_0|^{\frac{1}{2}}}
{|R_n|^{\frac{1}{2}}}
\]
with the following updated hyper-parameters:
\[
R_n=R_0 + X^{T}X
\]
\[
\beta_n=R_n^{-1}(R_0\beta_0 + X^{T}y)
\]
\[
a_{1n}=a_1+ \frac{n}{2}
\]
\[
a_{2n}=[\frac{ -\beta_n^{T}R_n\beta_n+y^{T}y+\beta_0^{T}R_0\beta_0} {2}
+ \frac{1}{a_2}]^{-1} .
\]

Details are for example in \citet{ohagan}. In the general genetic
model, the vector of parameters $\gamma$ is a linear transformation of
$\beta$, $\gamma=\omega\beta$, where the matrix $\omega$ is:
\[
\omega=\left[
\begin{array}{ccc}
1 & 0 & 0 \\
0 & 1 & 1 \\
0 & 2 & 4
\end{array}
\right] .
\]
Since $\gamma$ is a linear transformation of $\beta$, once a prior
distribution for $\beta$ is elicited, the prior distribution of $\gamma
$ is derived as:
\[
\gamma|\tau=\omega\beta|\tau\sim N( \omega\beta_0, \omega(\tau
R_0)^{-1}\omega^{T})
\]
while the prior for $\tau$ does not change with the
re-parameterization. If the prior distributions of the parameters
vectors are so defined, then it can be shown that $p(y|M_P )= p(y|M_G
)$ (see the Supplementary Material for details). In other words, the
marginal likelihood is invariant under linear transformations of the
regression coefficients.

Derivation of marginal likelihoods for additive, dominant, recessive,
and co-dominant models is different, as these models are defined by a
linear transformation of the parameters of the polynomial model and an
additional constraint. Formally, let $\alpha=\left[
\begin{array}{c}
\alpha_0 \\
\alpha_1
\end{array}
\right]$ denote the vector of parameters in any of these models. Then
we can define $\alpha=\left[
\begin{array}{c}
\alpha_0 \\
\alpha_1
\end{array}
\right] = \left[
\begin{array}{c}
\theta_0 \\
\theta_1
\end{array}
\right]| \theta_2=0$ where $\theta=\left[
\begin{array}{c}
\theta_0 \\
\theta_1 \\
\theta_2
\end{array}
\right]=\omega\beta$ and matrix $\omega$ depends on the specific
genetic model (see Table 2). If the vector $\beta$ follows a
multivariate normal distribution, $\theta$ also follows a multivariate
normal distribution, and so does the marginal distribution of $\theta
_2$ and $\alpha$ that is a conditional distribution. Starting from the
proper prior distributions for the vector of parameters $\beta$ and
precision $\tau$ priors, then proper prior distributions for $\alpha$
and $\tau$ are found to be:
\[
\tau\sim Gamma(a_1, a_2)
\]
\[
\alpha=\left[
\begin{array}{c}
\alpha_0 \\
\alpha_1
\end{array}
\right]=
\left[
\begin{array}{c}
\theta_0 \\
\theta_1
\end{array}
\right]| \theta_2=0
\sim N(\mu_0, \tau^{-1}\Sigma_0^{-1}) .
\]
$\mu_0$ and $\Sigma_0^{-1}$ can be obtained by using properties of the
conditional multivariate normal distribution (\citet*{eaton}) and are
summarized in Table 3. Using these derived priors, the marginal
likelihood for the additive, dominant, recessive, and co-dominant
models ($M_A$, $M_D$, $M_R$, and $M_C$, respectively) can be computed
in closed form. The derivation of the marginal likelihood for the
dominant model is detailed in the Supplementary Material. Derivation of
the marginal likelihood for the additive, recessive, and co-dominant
model is similar. Note that the derivation relies on the use of proper
prior distributions for the parameters of the polynomial model.

% matrix: b file: table1.tex 14 Sep 2001 18:06:44
%
%t2 ###
\begin{table}[htbp]
\small
\label{table2}
\begin{center}
\begin{tabular}{c|c|c|c|c|c}\hline
& 2-df General & Dominant & Recessive & Codomiant & Additive \\
& Model & Model & Model & Model & Model \\ \hline
$\omega$ &
$\left[
\begin{array}{ccc} 1 & 0 & 0 \\ 0 & 1 & 1 \\ 0 & 2 & 4
\end{array}
\right]$

&
$\left[
\begin{array}{ccc} 1 & 0 & 0 \\ 0 & 1 & 1 \\ 0 & 1 & 3
\end{array}
\right]$

&
$\left[
\begin{array}{ccc} 1 & 0 & 0 \\ 0 & 2 & 4 \\ 0 & 1 & 1
\end{array}
\right]$

&
$\left[
\begin{array}{ccc} 1 & 0 & 0 \\ 0 & 1 & 1 \\ 0 & 1 & 2
\end{array}
\right]$

&
$\left[
\begin{array}{ccc} 1 & 0 & 0 \\ 0 & 1 & 1 \\ 0 & 0 & 1
\end{array}
\right]$ \\ \hline
\end{tabular}
\caption{Specification of $\omega$ for five genetic models.}
\end{center}
\end{table}

Assuming that the 5 genetic models are a priori equally likely, the
Bayes rule to model selection is equivalent to choosing the genetic
model with the highest marginal likelihood or BF relative to the null
model (i.e. ratio of marginal likelihood of one of the five models and
the null model) (\citet{kass}). Once the most likely model is selected,
the parameter estimates of any of the five genetic models are the means
of the posterior distributions. The regression parameters in the
polynomial model are estimated by $\beta_n=R_n^{-1}(R_0\beta_0 +
X^{T}y)$ and using the one-to-one relationship, the parameters in the
general model can be estimated by $\gamma_n= \omega\beta_n$. The
relation between parameters of the polynomial models and the dominant,
recessive, co-dominant, and additive models can be used to derive their
posterior estimates. Specifically, from the set of relations:
\[
\beta|\tau\sim N(\beta_n, (\tau R_n)^{-1})
\]
\[
\theta=\omega\beta|\tau\sim N(\theta_n=\omega\beta_n, \omega(\tau
R_n)^{-1}\omega^{T})
\]
\[
\alpha=
\left[
\begin{array}{c}
\theta_0 \\
\theta_1
\end{array}
\right]| \theta_2=0
\sim N(\mu_n, \tau^{-1}\Sigma_n^{-1})
\]
and using the properties of the conditional multivariate normal
distribution, the point estimates $\mu_n$ for any model are found to be:
\[
\mu_n=
\left[
\begin{array}{c}
\theta_{n0} \\
\theta_{n1}
\end{array}
\right]+[S_{12}][S_{22}]^{-1}[0-\theta_{n2}]
\]
where $\omega(\tau R_n)^{-1}\omega^{T} = \tau^{-1}\left[
\begin{array}{cc}
S_{11} & S_{12} \\
S_{21} & S_{22}
\end{array}
\right]$, and $dim(S_{11})=2\times2$, $dim(S_{12})=2\times1$,
$dim(S_{21})=1\times2$, and $dim(S_{22})=1\times1$. Table 3 summarizes
the specification of ${\omega}$
and formulas for computing prior and updated hyper-parameters and
marginal likelihood for different genetic models discussed in this
section as well as the null model.

%% Table 2.

%%\begin{sidewaystable}
%
%t3 ###
\begin{table}[t!]
\def\arraystretch{1.3}
\tiny
%% \begin{center}
%
\begin{tabular}{p{0.6cm}|c|c|c}\hline
& Prior Hyper & Posterior Hyper & Marginal \\
& parameters & parameters & Likelihood\\ \hline

%poly
%
\begin{tabular}{c} $M_P$
\end{tabular}
&
\begin{tabular}{c}
$\beta_0 = [\beta_{00} \; \beta_{01} \; \beta_{02} ]^T $ \\
$R_0 = \left[
\begin{array}{ccc} r_{11} & r_{12} & r_{13} \\
r_{21} & r_{22} & r_{23} \\
r_{31} & r_{32} & r_{33}
\end{array}
\right] $ \\
$ a_1$ \\
$ a_2$
\end{tabular}
&
\begin{tabular}{c}
$\beta_n= R_n^{-1} (R_0\beta_0+X^T y)$ \\
$R_n = R_0 + X^Ty$ \\
$\alpha_{1n} = \alpha_1 +n/2$ \\
$\alpha_{2n} = \left( \frac{-\beta_n^T R_n \beta_n +y^T y }{2} \right.
$ \\
$ \left. \frac{\beta_0^T R_0 \beta_0}{2} + \frac{1}{\alpha_2} \right)^{-1}$
\end{tabular}
&
\begin{tabular}{c} $p(y|M_p) = \frac{1}{ \alpha_2^{\alpha_1} \Gamma
(\alpha_1) } $\\
$ \frac{1}{ (2 \pi)^{n/2} } \frac{ |R_0|^{1/2} }{|R_n|^{1/2}}$ \\
$ \alpha_{2n}^{\alpha_{1n}} \Gamma(\alpha_{1n})$
\end{tabular}
\\ \hline

%genog
%
\begin{tabular}{c} $M_G$
\end{tabular}
&
\begin{tabular}{c} $ \gamma_0= \omega\beta_0$ \\ $R_0' = \omega
(R_0)^{-1} \omega^T $ \\
$ a_1$ \\
$ a_2$
\end{tabular}
&
\begin{tabular}{c}
$ \gamma_n = \omega\beta_n$ \\
$R_n' = \omega(R_n)^{-1} \omega^T $\\
$\alpha_{1n} = \alpha_1 +n/2$ \\
$\alpha_{2n} = \left( \frac{-\beta_n^T R_n \beta_n +y^T y }{2} \right.
$ \\
$ \left. \frac{\beta_0^T R_0 \beta_0}{2} + \frac{1}{\alpha_2} \right)^{-1}$
\end{tabular}
&
\begin{tabular}{c}
$p(y|M_G) = \frac{1}{ \alpha_2^{\alpha_1} \Gamma(\alpha_1) } $\\
$ \frac{1}{ (2 \pi)^{n/2} } \frac{ |\omega(R_0)^{-1} \omega^T|^{1/2}
}{|(\omega^{-1})^T R_n \omega^{-1}|^{1/2}}$ \\
$ \alpha_{2n}^{\alpha_{1n}} \Gamma(\alpha_{1n})=p(y|M_p)$
\end{tabular}
\\ \hline

%dom
%
\begin{tabular}{c} $M_D$
\end{tabular}
&
\begin{tabular}{c} Let \\ $\omega\beta_0 = \theta_0 =
\left(
\begin{array}{c} \theta_{00} \\ \theta_{01} \\ \theta_{02}
\end{array}
\right)$
\end{tabular}
&
\begin{tabular}{c} Let \\ $\omega\beta_n = \theta_n =
\left(
\begin{array}{c} \theta_{n0} \\ \theta_{n1} \\ \theta_{n2}
\end{array}
\right)$
\end{tabular}
&
\begin{tabular}{c} $p(y|M_i) = \frac{1}{ \alpha_2^{\alpha_1} \Gamma
(\alpha_1) } $\\
$ \frac{1}{ (2 \pi)^{n/2} } \frac{ |\Sigma_0|^{1/2} }{|\Sigma
_n|^{1/2}}$ \\
$ \alpha_{2n}^{\alpha_{1n}} \Gamma(\alpha_{1n})$
\end{tabular}
\\

%rec
%
\begin{tabular}{c} $M_R$
\end{tabular}
&
\begin{tabular}{c} $\omega(\tau R_0)^{-1} \omega^T = S_0 =$ \\
$ \tau^{-1} \left(
\begin{array}{ccc} s_{0}^{11} & s_0^{12} & s_0^{13} \\ s_{0}^{21} &
s_0^{22} & s_0^{23} \\ s_{0}^{31} & s_0^{32} & s_0^{33}
\end{array}
\right)$
\end{tabular}
&
\begin{tabular}{c} $\omega(\tau R_n)^{-1} \omega^T = S_n =$ \\
$ \tau^{-1} \left(
\begin{array}{ccc} s_{n}^{11} & s_n^{12} & s_n^{13} \\ s_{n}^{21} &
s_n^{22} & s_n^{23} \\ s_{n}^{31} & s_n^{32} & s_n^{33}
\end{array}
\right)$
\end{tabular}
&
\begin{tabular}{c} where $i$= dominant, recessive \\ codominant or
additive
\end{tabular}
\\

%cod
%
\begin{tabular}{c}$M_C$
\end{tabular}
&
\begin{tabular}{c} $\mu_0 = \left(
\begin{array}{c} \theta_{00} \\ \theta_{01}
\end{array}
\right) -$ \\
$ \left(
\begin{array}{c} s_0^{13} \\
s_0^{23}
\end{array}
\right) (s_0^{33})^{-1} \theta_{02} $
\end{tabular}
&
\begin{tabular}{c} $\mu_n = \left(
\begin{array}{c} \theta_{n0} \\ \theta_{n1}
\end{array}
\right) -$ \\
$ \left(
\begin{array}{c} s_n^{13} \\
s_n^{23}
\end{array}
\right) (s_n^{33})^{-1} \theta_{n2} $
\end{tabular}
&
\\

%add
%
\begin{tabular}{c} $M_A$
\end{tabular}
&
\begin{tabular}{c} $\Sigma_0^{-1} = \left(
\begin{array}{cc} s_0^{11} & s_0^{12} \\ s_0^{21} & s_0^{22}
\end{array}
\right) -$ \\
$ \left(
\begin{array}{c} s_0^{13} \\
s_0^{23}
\end{array}
\right) (s_0^{33})^{-1} (s_0^{31} \; s_0^{32}) $
\end{tabular}
&
\begin{tabular}{c} $\Sigma_n^{-1} = \left(
\begin{array}{cc} s_n^{11} & s_n^{12} \\ s_n^{21} & s_n^{22}
\end{array}
\right) -$ \\
$ \left(
\begin{array}{c} s_n^{13} \\ s_n^{23}
\end{array}
\right) (s_n^{33})^{-1} (s_n^{31} \; s_n^{32}) $
\end{tabular}
&
\\

& $a_1$; $a_2$
&
\begin{tabular}{c} $a_{1n} = a_1 +n/2$ \\
$a_{2n} = \left( \frac{-\mu_n^T \Sigma_n \mu_n +y^T y }{2} \right. $ \\
$ \left. \frac{\mu_0^T \Sigma_0 \mu_0}{2} + \frac{1}{a_2} \right
)^{-1}$
\end{tabular}
& \\ \hline

%null
%
\begin{tabular}{c} $M_N$
\end{tabular}
&
\begin{tabular}{c}
$\beta_0 = [\beta_{00} ] $ \\
$R_0 = ( r_{11} ) $ \\
$ a_1$ \\
$ a_2$
\end{tabular}
&
\begin{tabular}{c}
$\beta_n= R_n^{-1} (R_0\beta_0+1^T y)$ \\
$R_n = R_0 + 1^Ty$ \\
$\alpha_{1n} = \alpha_1 +n/2$ \\
$\alpha_{2n} = \left( \frac{-\beta_n^T R_n \beta_n +y^T y }{2} \right.
$ \\
$ \left. \frac{\beta_0^T R_0 \beta_0}{2} + \frac{1}{\alpha_2} \right)^{-1}$
\end{tabular}
&
\begin{tabular}{c} $p(y|M_{null}) = \frac{1}{ \alpha_2^{\alpha_1} \Gamma
(\alpha_1) } $\\
$ \frac{1}{ (2 \pi)^{n/2} } \frac{ |R_0|^{1/2} }{|R_n|^{1/2}}$ \\
$ \alpha_{2n}^{\alpha_{1n}} \Gamma(\alpha_{1n})$
\end{tabular}
\\ \hline

\end{tabular}
\caption{Specification prior hyper-parameters, updated
hyper-parameters, and marginal likelihood for each model.
$M_P$=Polynomial Model, $M_G$=Genotypic Model, $M_D$=Dominant Model,
$M_R$=Recessive Model, $M_C$=Codominant Model, $M_A$=Additive Model,
and $M_N$=Null Model.}\label{table:two}
%% \end{center}
\end{table}
%
%%\end{sidewaystable}

\noindent
%s4 ###
\section{Simulation Studies}
Three simulation studies were conducted to assess false and true
positive rates of the Bayesian procedure with polynomial models and
compared to the frequentist approach in which the association with
minimum p-value is selected. Simulation study (1) was designed to
evaluate the false positive rates of the polynomial model approach for
various selection criteria. Real genotype data from two GWASs of
different sample sizes were used and the quantitative trait in each set
was randomly permuted to create data sets with no true positive
associations. Simulation study (2) was designed to compare sensitivity
and specificity of our proposed method and the standard approach by
simulating genetic data that mimic the GWAS setting with causal SNPs
(i.e. SNPs truly associated with the trait) having different modes of
inheritance, each SNP explaining the same proportion of the trait
variability. Simulation (3) modified the design of simulation (2) and
let SNPs explain varying proportions of the trait variability.

\noindent
%s4.1 ###
\subsection{Simulation Study (1)}
\underline{Data}: Two real datasets were used. The first data set
consisted of genotype data of 201 unrelated offspring of centenarians
from the New England Centenarian Study (NECS) (\url
{http://www.bumc.bu.edu/centenarian/}) (\citet{sebas2012}). The
genotype data were described in \citet{sebas.sig}. The quantitative
trait in this analysis was a neuroticism score measured in the NEO-Five
Factor Inventory (NEO-FFI), which is a 60-item (12 items per domain)
measure of five personality dimensions (neuroticism, extraversion,
openness, agreeableness, and conscientiousness) (\citet*{costa1992}).
Previous studies have shown that the estimated heritability of
neuroticism is approximately 25\% (\citet{baeneo,pilia}). The second
data set consisted of 843 unrelated African-American subjects with
sickle cell anemia enrolled in the Cooperative Study of Sickle Cell
Disease (CSSCD) (\texttt{\surl{https://biolincc.nhlbi.nih.\\gov/studies/csscd/}})
(\citet{gaston}). In this cohort, the trait is the percent of fetal
hemoglobin in the total hemoglobin. The percent fetal hemoglobin is a
major modulator of hematologic and clinical complications of sickle
cell anemia (\citet{akin}). Studies have shown that there is a strong
genetic basis of fetal hemoglobin and a well-established gene that
affects this trait is \textit{BCL11A} (\citet{bae2012}). The estimated
heritability ranges from 60.9\% to 89\% (\citet{garner,pilia}). Both
studies were approved by the institutional review board of each
participating institution, and standard quality control procedures were
performed on both genotype data (\citet{bae2012,sebas.sig}).

\noindent
\underline{Methods}: Initially 254,612 and 486,331 autosomal SNPs were
available for analysis in the two cohorts (NECS and CSSCD),
respectively. It is well known that SNPs in close proximity tend to be
correlated with each other (\citet{slatkin}), and this non-random
correlation can bias the assessment of false positive rates. In order
to avoid this problem, SNPs whose pairwise correlation was $r^2>0.2$
were removed using the PLINK software (\citet{purcell}). After this
pruning, 50,894 and 140,864 independent SNPs were left for analysis in
the NECS and CSSCD sample, respectively. In both sets, 50,000 SNPs were
randomly chosen from each set and 10,000 simulations were performed by
permuting the trait values at random. Two approaches were evaluated in
this simulation study: 1) the proposed method, in which the best
genetic model was selected based on the maximum BF for each SNP; and 2)
the frequentist approach, in which five genetic models were fitted and
the best model was selected based on the minimum p-value for each SNP.
For the genotypic model (2 degrees of freedom) in the frequentist
approach, the minimum of the two p-values was chosen. Various threshold
criteria for the two approaches were explored and the number of
significant associations detected for varying thresholds was recorded.
False positive rates were computed as the rates of significant associations.

%s4.2 ###
\subsection{Simulation Study (2)}
\underline{Data}: In order to assess the true positive rates of our
proposed method and the standard approach, genetic data were simulated
with known causal SNPs, each explaining the same proportion of the
trait variance. A modification of the simulation procedure described in
\citet{yip} was used to simulate the data but an additional source of
variability was introduced as well as SNPs with dominant and recessive
mode of inheritance, in addition to additive effects. Several scenarios
were considered by using different sample sizes (N=1,000, 10,000,
20,000, 50,000, and 100,000) and different heritability (\citet{ober})
of the quantitative traits ($h^2$=0.2, 0.4, and 0.6). Heritability is
defined as the proportion of the total variance of the trait that is
explained by the genetic effect and the higher the heritability the
larger the genetic contribution to the trait. A total of 500,000 SNPs
were simulated in each run and included 100 causal SNPs: 34 with
additive effects, 33 with dominant effects, and 33 with recessive
effects. We assumed that each causal SNP explained 1/100 of the total
genetic variance so that, for example, when the total heritability was
0.2 and 20\% of the total phenotypic variance was due to the genetic
variance, each causal SNP explained 1/100 of the total genetic variance
and hence the SNP-specific heritability was 0.002.

\noindent
\underline{Methods}: The following steps describe the simulation scheme.

\textit{Step 1. Generate minor allele frequency for each SNP}

The minor allele frequency (MAF: frequency of B allele) for each SNP
was randomly drawn from a Beta distribution $Beta(2,8)$, which
represents the distribution of commercially available chips (\citet
{yip}). We also used a standard quality control procedure by excluding
any SNPs with MAF less than 0.01.

\textit{Step 2. Generate the genotype}

Genotypes for each SNP (AA, AB, BB) were generated assuming
Hardy-Weinberg equilibrium. Essentially, if $p$ is the prevalence of
the A allele in the population, Hardy-Weinberg equilibrium (HWE) law
states that the prevalence of the three genotypes will be $p^2$,
$2p(1-p)$, $(1-p)^2$ (\citet{weinberg}). These expected genotype
frequencies were used to simulate the genotype data, given $p$.

\textit{Step 3. Select the causal SNPs}

100 causal SNPs from the total SNPs were randomly chosen and assigned
the mode of inheritance randomly to the selected SNPs.

\textit{Step 4. Determine the effect size for each causal SNP}

The effect size $a_j$ for each $j^{th}$ causal SNP (j=1, 2,{\ldots},
100) was computed from the formula:
\[
h_j^2=\frac{\sigma_{Add,j}^2 + \sigma_{Dom,j}^2}{\sigma_{Total}^2}
=\frac{2p_j(1-p_j)[a_j+d_j(1-2p_j)]^2+[2p_j(1-p_j)d_j]^2}
{\sigma_{Total}^2}
\]
where $\sigma_{Add,j}^2$ is the additive genetic variance of the
$j^{th}$ causal SNP, $\sigma_{Dom,j}^2$ is the dominance genetic
variance of the $j^{th}$ causal SNP, $\sigma_{Total}^2$ is the total
phenotypic variance, $p_j$ is the MAF for the $j^{th}$ causal SNP,
$a_j$ is the additive genetic effect at the $j^{th}$ causal SNP, and
$d_j$ is the dominance genetic effect at the $j^{th}$ causal SNP. The
parameter $h_j^2$ is the locus-specific heritability, which was assumed
to be $\frac{h^2}{100}$. This is the amount of heritability that is
contributed by the $j^{th}$ causal SNP and hence all causal SNPs
contribute to the total heritability by an equal amount. In the above
formula, note that the genetic variance is decomposed\vadjust{\goodbreak} into the additive
and dominance variance component. The additive genetic variance implies
that each additional copy of an allele contributes a fixed amount of
effect $a_j$ to the trait. Under this assumption, the trait value of
the heterozygote (AB) would be the midpoint between the two homozygotes
(AA and BB). On the other hand, when there exists dominance genetic
variance, the trait value of the heterozygote will deviate from the
midpoint between the two homozygotes, and the degree of deviation is
expressed by the quantity $d_j$. Therefore, it follows that $d_j=0$ for
any SNP with additive effect, and $d_j=a_j$ for any SNP with dominant
effect, and $d_j=-a_j$ for any SNP with recessive effect. We also
assumed $\sigma_{Total}^2=1$. Note that only the three genetic models
(additive, dominant, and recessive) were considered.

\textit{Step 5. Generate the phenotypic value based on the causal SNPs}

Let $y_i$ denote the phenotypic value for $i^{th}$ individual. For each
causal SNP, the SNP contribution to the trait was randomly generated as
\[
X_{ij} \sim N(a_j G_{ij}, \frac{\sigma_{Total}^2}{100})
\]
where $a_j$ is the effect of the $j^{th}$ causal SNP (computed in the
previous step) and $G_{ij}$ is genotype coding for the $i^{th}$
individual at the $j^{th}$ causal SNP that was generated in Step~2. For
an additive causal SNP, $G_{ij}$= number of minor allele (0, 1, 2). For
a dominant causal SNP, $G_{ij}=1$ if the genotype is AB or BB (0
otherwise). For a recessive causal SNP, $G_{ij}=1$ if the genotype is
BB (0 otherwise). Then, the phenotypic value is:
\[
y_i=\sum\limits_j X_{ij}
\]
and $E(y_i)=\sum\limits_j a_j G_{ij}$ and $Var(y_i)=\sigma_{Total}^2=1$.

\textit{Step 6. Perform association tests using our method and the
standard approach}

\textit{Step 7. Repeat 100 times}

In each simulated data set, the empirical false positive rates in the
two approaches were evaluated to determine thresholds for p-values and
BF with the same false positive rates. Specifically, in each simulated
set the number of false positive associations (significant associations
of null SNPs) in the frequentist results with p-values $p<1\times
10^{-7}$, $5\times10^{-7}$, $1\times10^{-6}$, and $5\times10^{-6}$ were
counted and the BF thresholds that produced the same number of false
positive associations in the Bayesian approach were detected. Using
these p-values and BF thresholds that produced the same empirical false
positive rates, the power of the two approaches was evaluated. Two
types of power were considered: (1) the number of causal SNPs detected
as associated regardless of whether the correct genetic model was
identified and (2) the number of causal SNPs detected with the true
genetic model.

%s4.3 ###
\subsection{Simulation Study (3)}
\underline{Data and Methods}: The limitation of Simulation Study (2) is
the assumption that each SNP accounts for the same proportion of the
trait variability. Therefore, the scheme of the Simulation Study
(2)\vadjust{\goodbreak}
was modified to let causal SNPs explain varying proportions of the
trait variability. In this modified design, the genetic variances of
dominant and recessive SNPs were increased, while decreasing the
genetic variance of additive SNPs. This trade-off was necessary to
maintain the same total heritability used in Simulation Study (2) for
proper comparison later. The following two cases were considered: 1)
when $h_j^2$ was halved for additive SNPs and 2) when $h_j^2$ was
quartered for additive SNPs. In case 1), this resulted in increasing
$h_j^2$ by 25\% for both dominant and recessive SNPs. In case 2), this
resulted in increasing $h_j^2$ by 37.5\% for both dominant and
recessive SNPs. Under these changes, the effect sizes were generated
based on Step 4 in the previous section and the rest of the simulation
design remained the same.

% matrix: b file: table1.tex 14 Sep 2001 18:06:44
%
%t4 ###
\begin{table}[htbp]
\small
\label{table3}
\begin{center}
\begin{tabular}{|p{1.2cm}|p{1.5cm}|p{1.5cm}|p{1.5cm}|p{1.5cm}|p{1.5cm}|p{1.5cm}|}
\multicolumn{7}{l}{a) Bayesian Polynomial Model Approach} \\ \hline
& BF$>$100 & BF$>$ 500 & BF$>$1000 &BF$>$1500 & BF$>$3000 & BF$>$5000
\\ \hline

NECS data & $9.0\times10^{-4}$ & $1.4\times10^{-4}$ & $8.0\times
10^{-5}$ & $4.0\times10^{-5}$ & $2.0\times10^{-5}$ &$0.0\times10^{-0}$
\\ \hline
CSSCD data & $6.8\times10^{-4}$ & $1.2\times10^{-4}$ & $6.0\times
10^{-5}$ & $4.0\times10^{-5}$ & $2.0\times10^{-5}$ &$0.0\times10^{-0}$
\\ \hline

\multicolumn{7}{l}{} \\
\multicolumn{7}{l}{b) Frequentist Approach} \\ \hline
& $p<10^{-3}$ & $p<10^{-4}$ &$p<10^{-5}$ &$p<10^{-6}$ &$p<10^{-7}$
&$p<10^{-8}$ \\ \hline

NECS data & $3.4\times10^{-3}$ & $4.0\times10^{-4}$ & $4.0\times
10^{-5}$ & $0.0\times10^{-0}$ & $0.0\times10^{-0}$ &$0.0\times10^{-0}$
\\ \hline
CSSCD data & $3.3\times10^{-3}$ & $3.8\times10^{-4}$ & $4.0\times
10^{-5}$ & $0.0\times10^{-0}$ & $0.0\times10^{-0}$ &$0.0\times10^{-0}$
\\ \hline
\end{tabular}
\caption{Median false positive rates in the NECS and CSSCD data in
10000 permutations (Simulation Study 1). BF=Bayes Factor; p=p-value.}
\end{center}
\end{table}

\noindent
\underline{Results}: Table 4 shows the median false positive rates at
varying significance thresholds in the two sets included in Simulation
Study (1). Setting the thresholds to maximum BF $>$ 1500 in our
approach and minimum p-value $<$ $10^{-5}$ in the standard approach
yields the same median false positive rate of $4\times10^{-5}$ in both
data sets. This translates into 2 false positive associations among
50,000 SNPs.

%f1 ###
\begin{figure}[h!]
\begin{center}
\includegraphics{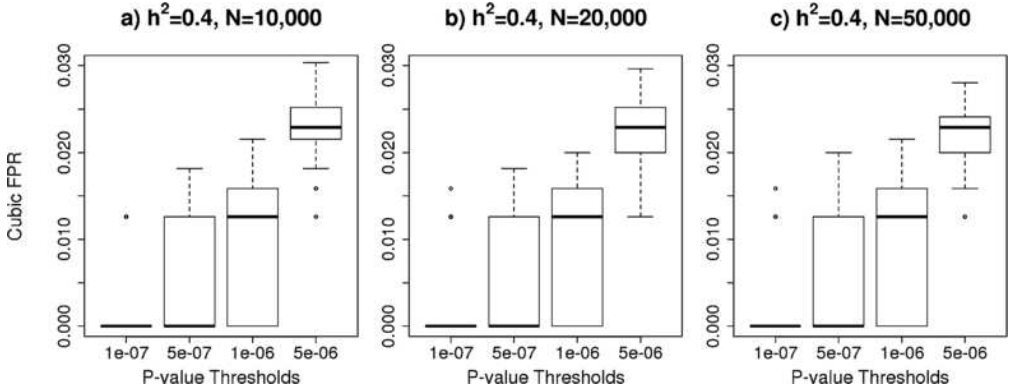}
\end{center}
\caption{Box plot of cubic root of empirical false positive rates
(y-axis) for different p-value thresholds (x axis), and increasing
sample sizes. The results are based on the simulation scenario when the
heritability was 0.4 and the sample sizes were 10,000, 20,000, and
50,000 (panel a, b, and c, respectively).}
\label{fig1}
\end{figure}
%
%f2 ###
\begin{figure}[h!]
\begin{center}
\includegraphics{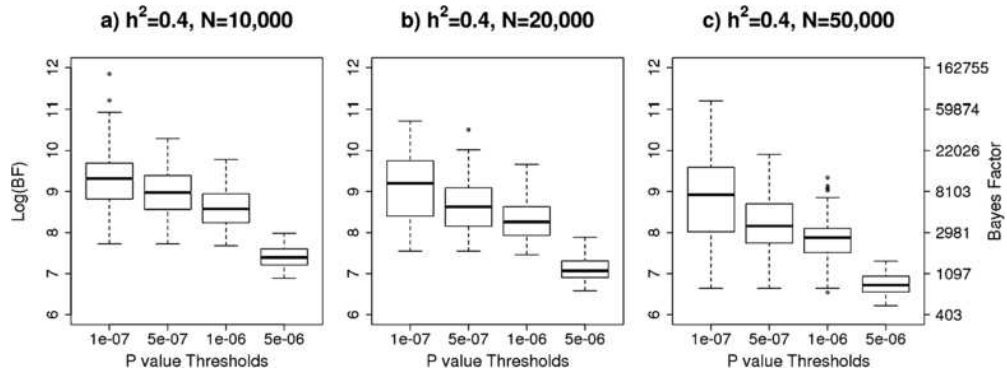}
\end{center}
\caption{Box plot of log-transformed BF (y-axis) for different p-values
thresholds (x-axis) and increasing sample sizes. The results are based
on the simulation scenario when the heritability estimate was 0.4 and
the sample sizes were 10,000, 20,000, and 50,000 (panel a, b, and c
respectively).}
\label{fig2}
\end{figure}
%
%f3 ###
\begin{figure}[h!]
\begin{center}
\includegraphics{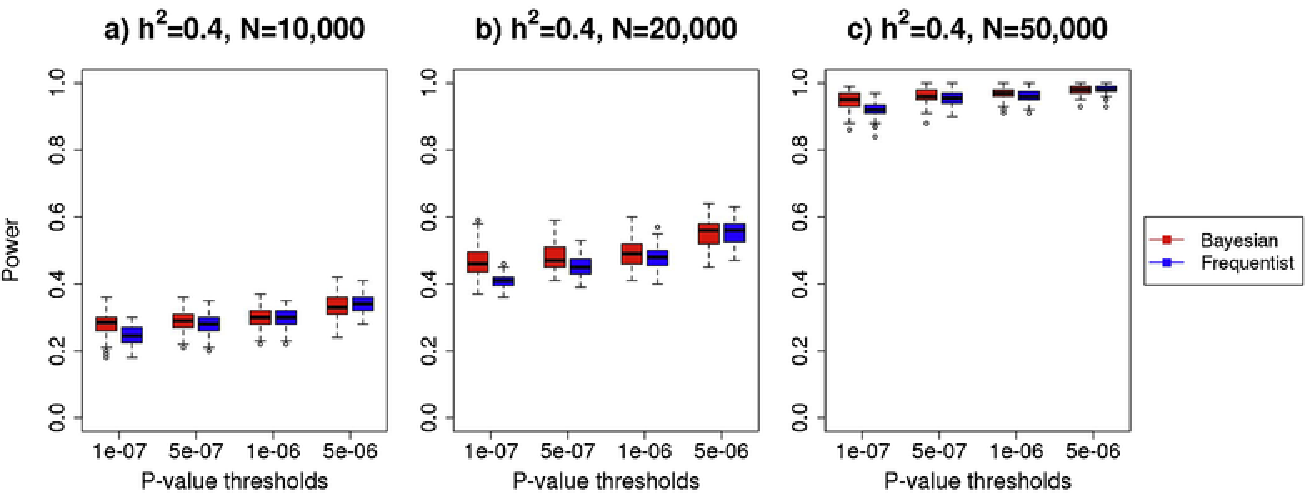}
\end{center}
\caption{Box plot of power of the two approaches at different
significance thresholds (Red: Bayesian approach; Blue: Frequentist
approach). The results are based on the simulation scenario when the
heritability estimate was 0.4 and the sample sizes were 10,000, 20,000,
and 50,000 (panel a, b, and c, respectively).}
\label{fig3}
\end{figure}

Figures 1-3 summarize the results of Simulation Study (2) for the
scenario in which the total heritability is 0.4 and the sample sizes
are 10,000, 20,000, and 50,000. The full set of results can be found in
the Supplementary Materials. With a sample of 1000, neither approach
detects any causal SNPs, while almost all causal SNPs are detected when
the sample size is 100,000, regardless of the heritability. Figure 1
shows the distribution of the empirical false positive rate for
different p-value thresholds and Figure 2 shows the distribution of the
BF that would produce the same empirical false positive rates of the
frequentist procedure. Figure 3 shows the box plot of the true positive
rate (proportion of detected causal SNPs) of the two approaches at
varying significance thresholds. Finally, Table 5 shows the mean number
of additive, dominant, and recessive SNPs that are correctly identified
(out of 34, 33, and 33, respectively) in the two approaches at
successive thresholds.

The mean and standard deviation of the quantitative traits were 2.55
and 1.10 when the total heritability was 0.4. Several points are
noteworthy. The first point is that, as expected, lower heritability of
the trait results in a smaller number of detections. Even when the
total heritability was relatively high ($h^2$=0.6), both approaches
detected about half of the causal SNPs with N=10,000. At the most
stringent significance threshold of $p<1\times10^{-7}$, the Bayesian
approach correctly identified 53.75 causal SNPs on average and the
frequentist approach correctly identified 46.52 causal SNPs on average
when heritability is 0.6 and N is 20,000 (see Supplementary Materials
for detail). This result is consistent with findings from other authors
that large sample sizes are needed to detect many casual variants that
explain a small proportion of variability. For example, in \citet{park}
authors have shown that they need approximately N=25,000 to detect 25
loci out of 201 causal variants with 80\% power for a highly heritable trait.

Secondly, we observed that the false positive rate of decision rules
based on BF decreases as the sample size increases, given a fixed BF
threshold. This property has also been noted in \citet{mathews} and
\citet{wakefield2012} and implies that we can relax the thresholds for
BF as we increase the sample size and better leverage the increased
sample size than frequentist procedures. Figure 2 illustrates this
property graphically. For example, at a fixed p-value threshold $1\times
10^{-7}$, the median BF threshold needed to obtain the same false
positive rate decreases from 11122 to 9866 to 7489 as the sample size
increases from 10,000 to 20,000 to 50,000. A similar pattern is
observed at all levels of false positive rates. In contrast, no such
pattern is observed in the frequentist approach, and the false positive
rates are invariant to sample sizes given a fixed p-value threshold in
the standard approach.

% matrix: b file: table1.tex 14 Sep 2001 18:06:44
%
%t5 ###
\begin{table}[t!]
\footnotesize
\label{table5}
\begin{center}
\begin{tabular}{p{1.6cm}cp{1.4cm}|cccc|cccc}  \hline
\multicolumn{3}{l}{} & \multicolumn{4}{c}{Bayesian} & \multicolumn
{4}{c}{Frequentist} \\ \hline
& N & Significance Threshold & A & D & R & Total & A & D & R & Total
\\ \hline
\multirow{12}{*}{\parbox{1.7cm}{Simulation Study (2) - Uniform
Contribution}} &

\multirow{4}{*}{10000} & $1\times10^{-7}$ &
15.6&0.7&0.9&17.2&17.3&0.3&0.3&17.9 \\
& & $5\times10^{-7}$ & 16.0&0.8&1.0&17.7&19.8&0.6&0.6&21.0 \\
& & $1\times10^{-6}$ & 16.3&0.9&1.2&18.4&20.8&0.7&0.8&22.3 \\
& & $5\times10^{-6}$ & 17.5&1.3&1.9&20.7&22.4&1.1&1.6&25.2 \\
\cline{2-11}

& \multirow{4}{*}{20000} & $1\times10^{-7}$ &
25.0&4.9&6.4&36.3&28.2&2.4&3.5&34.2 \\
& & $5\times10^{-7}$ & 25.0&5.6&7.5&38.0&28.2&4.1&5.4&37.7 \\
& & $1\times10^{-6}$ & 25.0&6.2&8.1&39.3&28.2&4.9&6.6&39.7 \\
& & $5\times10^{-6}$ & 25.0&8.8&11.0&44.8&28.3&7.5&10.1&45.9 \\
\cline{2-11}

& \multirow{4}{*}{50000} & $1\times10^{-7}$ &
30.0&27.1&30.5&87.6&31.2&23.2&28.6&82.9 \\
& & $5\times10^{-7}$ & 30.0&27.6&31.1&88.8&31.2&24.4&30.3&85.8 \\
& & $1\times10^{-6}$ & 30.0&27.8&31.4&89.3&31.2&24.6&31.0&86.7 \\
& & $5\times10^{-6}$ & 30.0&28.5&32.0&90.5&31.2&25.4&31.9&88.5 \\
\hline

\multirow{12}{*}{\parbox{1.7cm}{Simulation Study (3) - Case 1)}} &

\multirow{4}{*}{10000} & $1\times10^{-7}$ & 3.5&1.4&1.9&6.8&3.2&0.7&0.8&4.7
\\
& & $5\times10^{-7}$ & 3.8&1.5&2.1&7.4&4.9&1.0&1.4&7.3 \\
& & $1\times10^{-6}$ & 4.1&1.8&2.4&8.3&5.7&1.3&1.8&8.9 \\
& & $5\times10^{-6}$ &5.5&2.9&3.7&12.0&8.5&2.5&3.2&14.2 \\ \cline{2-11}

& \multirow{4}{*}{20000} & $1\times10^{-7}$ &
16.5&10.4&13.1&40.0&18.9&6.3&8.1&33.2 \\
& & $5\times10^{-7}$ & 17.0&11.5&14.5&43.1&21.2&9.0&11.5&41.7 \\
& & $1\times10^{-6}$ & 17.4&12.3&15.4&45.1&21.9&10.2&13.3&45.4 \\
& & $5\times10^{-6}$ & 18.5&14.7&18.7&51.9&23.6&12.9&17.7&54.2 \\
\cline{2-11}

& \multirow{4}{*}{50000} & $1\times10^{-7}$ &
26.3&29.4&32.7&88.4&29.0&26.6&32.2&87.8 \\
& & $5\times10^{-7}$ & 26.3&29.4&32.8&88.5&29.0&26.8&32.6&88.4 \\
& & $1\times10^{-6}$ & 26.3&29.5&32.8&88.7&29.0&26.9&32.7&88.5 \\
& & $5\times10^{-6}$ & 26.3&29.5&32.9&88.8&29.0&26.9&32.9&88.8 \\
\hline

\multirow{12}{*}{\parbox{1.7cm}{Simulation Study (3) - Case 2)}} &

\multirow{4}{*}{10000} & $1\times10^{-7}$ &
0.3&2.0&2.8&5.1&0.2&0.9&1.1&2.2 \\
& & $5\times10^{-7}$ & 0.3&2.3&3.1&5.7&0.5&1.7&2.1&4.2 \\
& & $1\times10^{-6}$ & 0.4&2.6&3.4&6.3&0.6&2.1&2.7&5.5 \\
& & $5\times10^{-6}$ & 0.7&3.9&5.0&9.6&1.2&3.5&4.5&9.2 \\ \cline{2-11}

& \multirow{4}{*}{20000} & $1\times10^{-7}$ &
3.9&13.5&16.8&34.2&3.4&8.8&11.2&23.5 \\
& & $5\times10^{-7}$ & 4.4&14.5&18.0&36.9&5.3&11.7&15.2&32.2 \\
& & $1\times10^{-6}$ & 4.8&15.2&19.0&39.0&6.4&12.7&16.9&35.9 \\
& & $5\times10^{-6}$ & 6.3&17.8&22.3&46.4&9.3&15.6&21.4&46.3 \\ \cline{2-11}

& \multirow{4}{*}{50000} & $1\times10^{-7}$ &
20.9&29.7&32.9&83.5&24.2&27.4&32.7&84.3 \\
& & $5\times10^{-7}$ & 21.3&29.7&32.9&84.0&25.3&27.5&32.9&85.6 \\
& & $1\times10^{-6}$ & 21.5&29.7&33.0&84.2&25.6&27.5&33.0&86.0\\
& & $5\times10^{-6}$ & 21.8&29.7&33.0&84.5&26.1&27.5&33.0&86.5 \\
\hline

\end{tabular}
\caption{Mean number of additive, dominant, and recessive SNPs
correctly identified (out of 34, 33, and 33, respectively) in the two
approaches when heritability is 0.4. A=additive; D=dominant; R=recessive.}
\end{center}
\end{table}

The third important point is that the Bayesian method we propose has a
slightly greater power for more stringent (lower) thresholds (see
Figure 3) than the frequentist approach. This result holds for all
sample sizes and all levels of heritability considered in the
simulations (see Supplement Figures S3 and S6). Also, at this stringent
threshold, the Bayesian approach recovered more correct genetic models
when the sample sizes were 20,000 and 50,000 (Table 5). Although our
method recovers less often than the frequentist approach SNPs with an
additive genetic effect, it identifies more often SNPs with a dominant
and recessive effect. When the sample size was 10,000, the Bayesian
approach recovered slightly fewer genetic models. In addition, both
approaches identified nearly 0 models that had either dominant or
recessive inheritance pattern when N=10,000 in simulation study (2). We
speculated that this may be due to the lack of power to detect
rare\vadjust{\goodbreak}
variants. For example, if we assume MAF=0.01, under HWE the expected
count of the homozygote group for the minor allele is only 1. As the
sample size increased, there was a substantial increase in
identification of dominant and recessive variants. Results from
simulation study (3) also support this conjecture. Increased effect
sizes for SNPs with dominant and recessive effects resulted in more
detection of these variants at the cost of loss of power for additive
SNPs. However, loss of power for additive SNPs was much greater than
increased power for dominant and recessive SNPs when the sample size
was 10,000. This result suggests that we need much higher sample sizes
to detect dominant and recessive variants, compared to the
additive~SNPs.

\noindent
%s5 ###
\section{Application to Real Data}
Using the thresholds that yielded the same false positive rate in the
two methods (maximum BF$>$1500 and minimum $p<1\times10^{-5}$) in
simulation study (1), we compared the results obtained with the two
methods in the cohorts described in the earlier section, using the SNP
sets generated after pruning the dependent SNPs. In the NECS data, out
of 50,894 tested SNPs, nine SNPs were found associated with neuroticism
using the polynomial model approach, whereas only five SNPs were
significant using the standard approach (Table 6). Four SNPs were
common in both analyses. For these four SNPs, the genetic models
selected agreed in the two approaches. This result suggests that the
Bayesian model selection procedures work well in the case of small
sample sizes and can potentially discover more variants.

In the CSSCD data, out of 140,864 tested SNPs, ten SNPs were associated
with fetal hemoglobin in both approaches, and eight SNPs were common in
both (Table 6). For these eight SNPs in common, five of them agreed in
the genetic model selection between the two approaches, but three SNPs
(rs2239580, rs12469604, and rs2034614) had discrepant results. For
rs2239580, the Bayesian procedures selected the dominant model, while
the standard approach identified the genotypic model. For rs12469604,
the Bayesian procedure selected the dominant model, while the additive
model had the minimum p-value. For rs2034614, the co-dominant model had
the maximum BF and the genotypic model had the minimum p-value.

Using our Bayesian polynomial model approach in the NECS data, 4 SNPs
had dominant models, 3 SNPs had additive models, 1 SNP had a
co-dominant model and 1 SNP had a recessive model. In the CSSCD data, 3
SNPs had dominant models, 3 SNPs had co-dominant models, 3 SNPs had
recessive models, and 1 SNP had an additive model. These results
suggest that different variants may influence the trait through
different genetic models. Some of these associations would not have
been captured if an additive model alone was used, and this highlights
the need to examine all possible genetic models in a computationally
efficient manner to ensure that we do not miss any interesting findings.

% matrix: b file: table1.tex 14 Sep 2001 18:06:44
%
%t6 ###
\begin{table}[htbp]
\scriptsize
\label{table6}
\begin{center}
\begin{tabular}{c|c|c|c|c|c|c|c|c}

\multicolumn{9}{l}{a) NECS Data} \\ \hline
\multicolumn{2}{l}{} & \multicolumn{7}{c}{Bayesian} \\ \hline
SNP &Chr/Gene & \multicolumn{2}{c|}{$BF_G$} & $BF_A$ & $BF_D$ & $BF_R$
& $BF_C$ & $BF_{max}$ \\ \hline

\rowcolor{Gray}rs850610&1/C1orf203& \multicolumn
{2}{c|}{$1.0e{5}$}&$8.0e{4}$&$1.2e{6}$&$1.5e{1}$&$1.5e{3}$&$1.2e{6}$ \\
\hline
\rowcolor{Gray}rs7666974&4/unknown& \multicolumn
{2}{c|}{$7.6e{2}$}&$1.4e{-1}$&$7.4e{1}$&$7.5e{1}$&$1.2e{4}$&$1.2e{4}$
\\ \hline
\rowcolor{Gray}rs2333166&4/unknown&\multicolumn
{2}{c|}{$4.6e{2}$}&$3.9e{2}$&$3.4e{3}$&$3.8e{-1}$&$1.5e{3}$&$3.4e{3}$
\\ \hline
\rowcolor{Gray}rs2801185&1/ESRRG&\multicolumn
{2}{c|}{$2.2e{2}$}&$4.4e{1}$&$4.9e{-1}$&$2.0e{3}$&$4.0e{-1}$&$2.0e{3}$
\\ \hline
rs1869676&8/unknown&\multicolumn
{2}{c|}{$1.4e{3}$}&$5.0e{3}$&$1.0e{1}$&$2.0e{3}$&$3.7e{1}$&$5.0e{3}$ \\
\hline
rs8064944&17/unknown&\multicolumn
{2}{c|}{$2.2e{2}$}&$1.0e{1}$&$4.3e{3}$&$2.9e{-1}$&$4.4e{-1}$&$4.3e{3}$
\\ \hline
rs3746314&19/C19orf12&\multicolumn
{2}{c|}{$1.8e{3}$}&$2.4e{3}$&$9.4e{2}$&$3.7e{1}$&$9.9e{-1}$&$2.4e{3}$
\\ \hline
rs9555139&13/unknown&\multicolumn
{2}{c|}{$6.5e{2}$}&$2.2e{3}$&$6.7e{1}$&$1.5e{2}$&$1.5e{1}$&$2.2e{3}$ \\
\hline
rs12770017&10/unknown&\multicolumn
{2}{c|}{$1.4e{2}$}&$3.0e{-1}$&$1.9e{3}$&$1.2e{-1}$&$1.9e{1}$&$1.9e{3}$
\\ \hline
rs1530239&2/IKZF2&\multicolumn
{2}{c|}{$7.9e{1}$}&$5.8e{1}$&$7.0e{-1}$&$2.1e{1}$&$5.9e{1}$&$5.9e{1}$
\\ \hline

\multicolumn{2}{l}{} & \multicolumn{7}{c}{Frequentist} \\ \hline
SNP &Chr/Gene & $P_{Het}$ & $P_{Hom}$ & $P_A$ & $P_D$ & $P_R$ & $P_C$ &
$P_{min}$ \\ \hline

\rowcolor
{Gray}rs850610&1/C1orf203&$2.0e{-7}$&$2.4e{-4}$&$2.5e{-7}$&$1.9e{-8}$&$3.3e{-2}$&$2.1e{-5}$&$1.9e{-8}$
\\ \hline
\rowcolor
{Gray}rs7666974&4/unknown&$8.4e{-4}$&$7.7e{-1}$&$1.5e{-1}$&$5.3e{-2}$&$6.9e{-4}$&$1.9e{-6}$&$1.9e{-6}$
\\ \hline
\rowcolor
{Gray}rs2333166&4/unknown&$5.8e{-6}$&$2.3e{-1}$&$5.5e{-5}$&$5.5e{-6}$&$4.4e{-1}$&$8.7e{-6}$&$5.5e{-6}$
\\ \hline
\rowcolor
{Gray}rs2801185&1/ESRRG&$7.3e{-1}$&$3.1e{-6}$&$1.9e{-3}$&$1.0e{-1}$&$2.8e{-6}$&$6.2e{-1}$&$2.8e{-6}$
\\ \hline
rs1869676&8/unknown&$8.1e{-1}$&$1.5e{-1}$&$3.5e{-4}$&$2.4e{-1}$&$1.1e{-4}$&$3.2e{-4}$&$1.1e{-4}$
\\ \hline
rs8064944&17/unknown&$4.0e{-4}$&$2.7e{-4}$&$2.2e{-2}$&$2.1e{-4}$&$2.8e{-1}$&$5.2e{-1}$&$2.1e{-4}$
\\ \hline
rs3746314&19/C19orf12&$2.2e{-2}$&$1.5e{-3}$&$8.8e{-4}$&$2.2e{-3}$&$6.0e{-3}$&$1.0e{-1}$&$8.8e{-4}$
\\ \hline
rs9555139&13/unknown&$4.2e{-1}$&$6.2e{-2}$&$8.2e{-3}$&$7.0e{-2}$&$1.2e{-2}$&$6.1e{-2}$&$8.2e{-3}$
\\ \hline
rs12770017&10/unknown&$2.6e{-4}$&$1.1e{-3}$&$9.6e{-1}$&$7.9e{-4}$&$3.8e{-1}$&$6.0e{-2}$&$2.6e{-4}$
\\ \hline
rs1530239&2/IKZF2&$3.5e{-2}$&$1.4e{-3}$&$3.7e{-6}$&$3.1e{-3}$&$2.3e{-5}$&$5.0e{-4}$&$3.7e{-6}$
\\ \hline

\multicolumn{9}{l}{} \\
\multicolumn{9}{l}{b) CSSCD Data} \\ \hline
\multicolumn{2}{l}{} & \multicolumn{7}{c}{Bayesian} \\ \hline
SNP &Chr/Gene & \multicolumn{2}{c|}{$BF_G$} & $BF_A$ & $BF_D$ & $BF_R$
& $BF_C$ & $BF_{max}$ \\ \hline

\rowcolor{Gray}rs6709302&2/BCL11A&\multicolumn
{2}{c|}{$1.7e{4}$}&$1.9e{5}$&$7.4e{3}$&$3.8e{2}$&$2.1e{1}$&$1.9e{5}$ \\
\hline
\rowcolor{Gray}rs7631659&3/unknown&\multicolumn
{2}{c|}{$1.2e{4}$}&$3.4e{1}$&$4.4e{-1}$&$1.6e{5}$&$2.2e{-1}$&$1.6e{5}$
\\ \hline
\rowcolor{Gray}rs13043968&20/unknown&\multicolumn
{2}{c|}{$1.9e{3}$}&$1.2e{1}$&$2.5e{-1}$&$3.1e{4}$&$2.1e{-1}$&$3.1e{4}$\\
\hline
\rowcolor{Gray}rs2239580&14/COCH&\multicolumn
{2}{c|}{$2.5e{3}$}&$1.1e{3}$&$3.0e{4}$&$1.5e{-1}$&$2.8e{4}$&$3.0e{4}$
\\ \hline
\rowcolor{Gray}rs6932510&6/RPS6KA2&\multicolumn
{2}{c|}{$5.9e{2}$}&$2.1e{3}$&$5.9e{3}$&$3.1e{-1}$&$3.7e{3}$&$5.9e{3}$\\
\hline
\rowcolor{Gray}rs1890911&14/unknown&\multicolumn
{2}{c|}{$2.5e{2}$}&$8.5e{-1}$&$1.5e{-1}$&$4.4e{3}$&$3.3e{-1}$&$4.4e{3}$\\
\hline
\rowcolor{Gray}rs12469604&2/unknown&\multicolumn
{2}{c|}{$6.1e{2}$}&$2.2e{3}$&$2.9e{3}$&$6.3e{-1}$&$1.8e{3}$&$2.9e{3}$
\\ \hline
\rowcolor{Gray}rs2034614&12/PRICKLE1&\multicolumn
{2}{c|}{$1.8e{2}$}&$1.5e{2}$&$1.7e{3}$&$1.6e{-1}$&$2.8e{3}$&$2.8e{3}$\\
\hline
rs2301819&4/TBC1D14&\multicolumn
{2}{c|}{$8.4e{1}$}&$5.0e{1}$&$1.9e{2}$&$1.9e{-1}$&$2.0e{3}$&$2.0e{3}$\\
\hline
rs9642124&7/unknown&\multicolumn
{2}{c|}{$3.8e{1}$}&$4.8e{-2}$&$3.2e{1}$&$9.2e{-1}$&$1.8e{3}$&$1.8e{3}$\\
\hline
rs11794652&9/FUBP3&\multicolumn
{2}{c|}{$9.9e{1}$}&$4.2e{2}$&$5.2e{1}$&$6.0e{2}$&$2.1e{-1}$&$6.0e{2}$\\
\hline
rs7975463&12/unknown&\multicolumn
{2}{c|}{$3.5e{1}$}&$1.3e{1}$&$1.8e{2}$&$8.4e{-2}$&$8.9e{2}$&$8.9e{2}$\\
\hline

\multicolumn{2}{l}{} & \multicolumn{7}{c}{Frequentist} \\ \hline
SNP &Chr/Gene & $P_{Het}$ & $P_{Hom}$ & $P_A$ & $P_D$ & $P_R$ & $P_C$ &
$P_{min}$ \\ \hline

\rowcolor
{Gray}rs6709302&2/BCL11A&$1.2e{-4}$&$2.5e{-7}$&$1.3e{-8}$&$8.0e{-7}$&$2.6e{-5}$&$1.5e{-2}$&$1.3e{-8}$\\
\hline
\rowcolor
{Gray}rs7631659&3/unknown&$9.1e{-1}$&$7.6e{-8}$&$7.0e{-3}$&$1.5e{-1}$&$7.5e{-8}$&$9.2e{-1}$&$7.5e{-8}$\\
\hline
\rowcolor
{Gray}rs13043968&20/unknown&$8.9e{-1}$&$5.4e{-7}$&$2.1e{-2}$&$2.8e{-1}$&$4.9e{-7}$&$6.7e{-1}$&$4.9e{-7}$\\
\hline
\rowcolor
{Gray}rs2239580&14/COCH&$1.7e{-7}$&$2.1e{-1}$&$4.9e{-6}$&$2.3e{-7}$&$5.7e{-1}$&$3.2e{-7}$&$1.7e{-7}$\\
\hline
\rowcolor
{Gray}rs6932510&6/RPS6KA2&$4.9e{-6}$&$2.5e{-1}$&$6.4e{-6}$&$3.3e{-6}$&$3.9e{-1}$&$6.5e{-6}$&$3.3e{-6}$\\
\hline
\rowcolor
{Gray}rs1890911&14/unknown&$6.7e{-1}$&$4.7e{-6}$&$1.5e{-2}$&$3.7e{-1}$&$2.6e{-6}$&$2.5e{-1}$&$2.6e{-6}$\\
\hline
\rowcolor
{Gray}rs12469604&2/unknown&$1.2e{-5}$&$2.3e{-1}$&$6.6e{-6}$&$7.2e{-6}$&$2.8e{-1}$&$1.4e{-5}$&$6.6e{-6}$\\
\hline
\rowcolor
{Gray}rs2034614&12/PRICKLE1&$7.7e{-6}$&$5.6e{-1}$&$9.4e{-5}$&$1.2e{-5}$&$8.7e{-1}$&$9.0e{-6}$&$7.7e{-6}$\\
\hline
rs2301819&4/TBC1D14&$1.9e{-5}$&$7.0e{-1}$&$3.5e{-3}$&$1.3e{-4}$&$3.7e{-1}$&$1.3e{-5}$&$1.3e{-5}$\\
\hline
rs9642124&7/unknown&$4.0e{-4}$&$8.7e{-1}$&$9.8e{-1}$&$1.6e{-2}$&$1.4e{-2}$&$1.6e{-5}$&$1.6e{-5}$\\
\hline
rs11794652&9/FUBP3&$9.9e{-2}$&$3.3e{-6}$&$7.9e{-6}$&$2.8e{-3}$&$1.1e{-5}$&$4.8e{-1}$&$3.3e{-6}$\\
\hline
rs7975463&12/unknown&$7.6e{-6}$&$2.8e{-1}$&$6.5e{-3}$&$4.6e{-5}$&$6.2e{-1}$&$1.2e{-5}$&$7.6e{-6}$\\
\hline

\end{tabular}
\caption{Significant results using two approaches in the a) NECS and b)
CSSCD data. BF=Bayes Factor ;P=p-value; G=genotypic; A=additive;
D=dominant; R=recessive; C=co-dominant; Het=heterozygote genotype
factor in the genotypic model; Hom=homozygote genotype factor in the
genotypic model. SNPs that were significant in both approaches are
highlighted in gray.
}
\end{center}
\end{table}

\noindent
%s6 ###
\section{Conclusion}
We propose a Bayesian approach to simultaneously detect the SNPs
associated with a continuous trait and the mode of inheritance. Our
Bayesian approach uses a polynomial parameterization of the SNP dosage
that can simultaneously represent different genetic models and a
coherent framework for model selection based on comparing different
models by their posterior probability
(\citet{wellcome,marchini,servin,guan,wakefield2008,newcombe,stephens,clark,bermejo,maller,xu}).
Crucial to our approach is the use of proper prior
distributions on the parameters of the polynomial model, from which the
prior distributions of specific genetic models can be derived. In
contrast to this coherent Bayesian approach, it is important to
emphasize that the frequentist approach does not have a clear way to
compare the genotypic model (2 degrees of freedom) to the other four
specific genetic models (1 degree of freedom). The evaluation of the
method in simulated data shows that the Bayesian method we propose has
a slightly higher power when we limit to false positive rates at very
small values and this is a particularly attractive property in
genome-wide association studies in which the large number of SNPs
analyzed requires to set the false positive rate to extremely small
numbers. An additional attractive feature of this method is the gain in
computations: The proposed method codes five genetic models
simultaneously using a single polynomial parameterization instead of
fitting five different genetic models for each SNP. This contrasts with
the standard approach in which recoding of the SNP genotype and
conducting 5 analyses is necessary to evaluate all five models
concurrently.\looseness=1

An important theoretical implication of this particular
parameterization is that it shows that different genetic models are
functionally related. We have shown that there is a mathematical
relationship between the parameters of the polynomial model and each of
the five genetic models. This relation also suggests that when all five
genetic models are evaluated the effective number of tests per SNP is
less than 5. In practice, GWASs suffer from severe correction for
multiple testing, and evaluation of several genetic models for each SNP
aggravates this issue. However, our work suggests that the correction
for multiple testing should be less severe as the effective number of
tests is less than the number of models fitted, when evaluating all
five genetic models, although it is not immediately obvious how
Bonferroni type corrections should benefit from this result.

The proposed work can be particularly useful for genome-wide data
consisting of millions of SNPs. This work, at the current state, is
limited to the case where the trait is quantitative, as we can obtain
closed form solutions for the marginal likelihood and BF. More work is
needed to evaluate a similar approach when the trait of interest is
binary or time-to-event. In particular, when performing logistic
regression in the GWAS context, alternative measures of associations
such as approximate Bayes Factor or Bayesian false-discovery
probability (\citet{wakefield2007,wakefield2008,wakefield2009}) can
be considered.\looseness=1

%% Supplement Material %%
% \sname{}\label{}
% \stitle{}
% \slink[url]{}
% \sdescription{}
\begin{supplement}
 \stitle{Supplementary Materials: Bayesian Polynomial Regression Models to Fit Multiple Genetic Models for Quantitative Traits}
 \sdatatype{.pdf}
 \sfilename{Supplementary Materials BA-Apr-22-2014.pdf}
 \slink[doi]{10.1214/14-BA880SUPP}
\end{supplement}

\bibliographystyle{ba}

\begin{acknowledgement}
This work was funded by the National Institute on Aging (NIA
U19-AG023122 to T.P.), the National Heart Lung Blood Institute
(R21HL114237 to P.S.), and the National Institute of Health (R01HL87681
to M.H.S.).
\end{acknowledgement}

\end{document}